\documentclass[conference]{IEEEtran}

\usepackage{graphicx}
\usepackage{moreverb}
\usepackage[colorlinks=true,citecolor=red]{hyperref}
\usepackage{graphicx}
\usepackage{float}
\usepackage{amssymb}
\usepackage{amsfonts}
\usepackage{multicol}
\usepackage{amsmath}
\usepackage{mathtools}
\usepackage[mathscr]{euscript}
\usepackage{booktabs}
\usepackage{makecell}
\usepackage{multirow}
\usepackage{amsbsy} 
\usepackage[hypcap=false, font=small]{caption}
\usepackage[noend,ruled,linesnumbered]{algorithm2e}
\usepackage{cite}
\usepackage{xfrac}
\usepackage{bbm}
\usepackage{color, colortbl}
\definecolor{Gray}{gray}{0.9}
\usepackage[first=0,last=9]{lcg}

\usepackage[absolute,showboxes]{textpos}
\PassOptionsToPackage{draft}{hyperref}
\usepackage{subfig}

\TPMargin{5pt}

\newcommand{\copyrightstatement}{
    \begin{textblock}{0.84}(0.08,0.01)    
         \noindent
         \footnotesize
         \copyright 2018 IEEE. Personal use of this material is permitted. Permission from IEEE must be obtained for all other uses, in any current or future media, including reprinting/republishing this material for advertising or promotional purposes, creating new collective works, for resale or redistribution to servers or lists, or reuse of any copyrighted component of this work in other works.
    \end{textblock}
}

\begin{document}

\copyrightstatement

\title{Power-Aware Virtual Network Function Placement and Routing using an Abstraction Technique\vspace{-2.9ex}}

\author{\IEEEauthorblockN{Amir Varasteh$^\dagger$, Marilet De Andrade$^\star$, Carmen Mas Machuca$^\dagger$, Lena Wosinska$^\star$, and Wolfgang Kellerer$^\dagger$ \\
$^\dagger$ Chair of Communication Networks, Department of Electrical and Computer Engineering,\\ Technical University of Munich, Germany\\
Email: \{amir.varasteh, cmas, wolfgang.kellerer\}@tum.de\\
$^\star$ Optical Networks Lab., KTH Royal Institute of Technology, Stockholm, Sweden\\ 
Email: \{marilet, wosinska\}@kth.se}\vspace{-3.3ex}
}
\maketitle

\begin{abstract}
The Network Function Virtualization (NFV) is very promising for efficient provisioning of network services and is attracting a lot of attention. NFV can be implemented in commercial off-the-shelf servers or Physical Machines (PMs), and many network services can be offered as a sequence of Virtual Network Functions (VNFs), known as VNF chains. Furthermore, many existing network devices (e.g., switches) and collocated PMs are underutilized or over-provisioned, resulting in low power-efficiency. In order to achieve more energy efficient systems, this work aims at designing the placement of VNFs such that the total power consumption in network nodes and PMs is minimized, while meeting the delay and capacity requirements of the foreseen demands. Based on existing switch and PM power models, we propose a Integer Linear Programming (ILP) formulation to find the optimal solution. We also propose a heuristic based on the concept of Blocking Islands (BI), and a baseline heuristic based on the Betweenness Centrality (BC) property of the graph. Both heuristics and the ILP solutions have been compared in terms of total power consumption, delay, demands acceptance rate, and computation time. Our simulation results suggest that BI-based heuristic is superior compared with the BC-based heuristic, and very close to the optimal solution obtained from the ILP in terms of total power consumption and demands acceptance rate. Compared to the ILP, the proposed BI-based heuristic is significantly faster and results in 22\% lower end-to-end delay, with a penalty of consuming 6\% more power in average.
\end{abstract}
\IEEEpeerreviewmaketitle
\vspace{-0.2cm}
\section{Introduction} \vspace{.13cm}
Network Function Virtualization (NFV) aims to tackle the limitations of hardware network functions \cite{halpern2015rfc}\cite{portal2012network}. In this platform, network function implementations have evolved by running their software over virtualized general purpose hardware. These functions are known as Virtual Network Functions (VNFs). In the NFV architecture, a commercial off-the-shelf Physical Machine (PM) can host several Virtual Machines (VMs), each VM implementing a network function with special software programs. NFV can bring many benefits to the telecommunication networks: openness of platforms, scalability and flexibility, performance improvement, and also cost reductions \cite{portal2012network}.\par
In modern telecommunication, with the growth of cloud computing and virtualization technologies, network providers start deploying their network services using VNFs. A service is usually composed of various VNFs (e.g., firewall, WAN optimizer, Network translation service) based on customer demands. This ordered sequence of VNFs form a service function (VNF) chain \cite{halpern2015rfc}. However, effective deployment and resource allocation of VNF chains is a complex, yet important challenge to overcome. Especially, power consumption of these networks has always been a crucial issue \cite{abts2010energy}. Bolla \textit{et al.} \cite{bolla2012cutting} show the aggressive trend of power consumption increase in the networks operated by the major telecom operators worldwide (e.g., AT\&T, Verizon). The high power consumption of these networks along with their high rate of growth and total carbon footprint have made it inevitable to apply Green Computing techniques and reduce the networks power consumption and guarantee the overall network scalability and sustainability.\par
Many studies have shown a massive waste of power consumed by idle or under-utilized devices. On one hand, PMs that are utilized by VNF instances are not power-proportional, since they consume more than 50\% of their maximum power when they are in idle state \cite{gandhi2009optimal}. Therefore, to provide power-efficiency, one solution is to increase PMs resource utilization, which leads to using lower number of active PMs. On the other hand, network devices are also an undeniable power-consuming part of telecommunication networks \cite{mahadevan2011energy}. Similar to the computation resources, network resources are also usually over-provisioned to support the maximum traffic. However, their utilization rarely reaches the peak network capacity \cite{heller2010elastictree}. Consequently, idle networking devices are not power-efficient, since an idle network switch can consume up to 90\% of the peak power consumption \cite{mahadevan2011energy}. Thus, minimizing the number of active network devices can also contribute to the overall power-efficiency.\par
Keeping in mind the above observations, in this work, we study the joint power-aware VNF chain placement and routing problem. The main goal is to minimize the amount of computation (i.e., PMs) and network (i.e., switches and links) resources required to realize VNF chains while meeting end-to-end delay requirements. To do so, we proposed the optimal power-aware placement of VNFs using an Integer Linear Programming (ILP) formulation. Due to the complexity of the ILP, we also proposed an efficient heuristic based on the Blocking Island (BI) resource abstraction paradigm. \par
The main contributions of this paper are: \textit{(i)} a solution to the VNF chain placement problem using ILP \textit{(ii)} Design of an efficient heuristic for joint power-aware VNF chain placement and routing using the BI resource abstraction technique, and \textit{(iii)} a new heuristic based on Betweenness Centrality (BC) property of graph \cite{freeman1977set}, which is used as baseline. 
The rest of the paper is organized as follows: Section~\ref{relatedwork} reviews the related works. Section~\ref{systemmodel} presents the system model and problem formulation. Then, the two proposed heuristics are introduced in Section~\ref{solution}, followed by the performance evaluation in Section~\ref{evaluation}. Finally, Section~\ref{conc} concludes the paper.
\section{Related Work} \label{relatedwork} \vspace{.15cm}
A comparison of different power aware placement solutions with respect to the proposed one has been depicted in Table \ref{table1}. Power-aware VM placement has been widely studied in the cloud computing environment \cite{yao2016joint, jin2013joint, dalvandi2017application, mosa2016optimizing}. Surveys on VM placement techniques can be found in \cite{lopez2015virtual, varasteh2017server}. Most of the VM placement studies are focused on minimizing either PM or network power consumption. However, VNF chain placement strategies differ from VM placement solutions in mainly ways. Firstly, the VNF chain placement problem, considers VNF chains, which requires a specific ordered set of VNFs, which is not the case in VM placement problems. Additionally, VNF chain placement problem can be considered as two NP-hard problems: Virtual Network Embedding, and Location-Routing Problem \cite{hmaity2016virtual}, which is not the case for VM placement. Thus, these specific characteristics and constraints has made the VNF placement problem more complex. A comprehensive survey on VNF chain placement is presented in \cite{bhamare2016survey}.
\begin{table}[t]
\centering
\setlength\belowcaptionskip{12pt}
\caption{Characteristics considered in the related work}
\label{table1}
\resizebox{\columnwidth}{!}{
\begin{tabular}{cccccc}
\toprule
\textbf{Works} & \textbf{PM Power} & \textbf{Network Power}  & \textbf{Delay} & \textbf{Functions Order} & \textbf{Instance Sharing} \\ \hline 
\cite{mosa2016optimizing}\cite{huang2013energy} & $\checkmark$ & $\times$ & $\times$ & $\times$ & $\times$  \\ \hline
\cite{yao2016joint}\cite{jin2013joint} & $\checkmark$ & $\checkmark$ & $\times$ & $\times$ & $\times$  \\ \hline
\cite{dalvandi2017application} & $\checkmark$ & $\times$ & $\checkmark$ & $\times$ & $\times$  \\ \hline
\cite{fang2013vmplanner} & $\times$ & $\checkmark$ & $\times$ & $\times$ & $\times$  \\ \hline
\cite{kim2017vnf}\cite{pham2017traffic} & $\checkmark$ & $\times$ & $\checkmark$ & $\checkmark$ & $\checkmark$  \\ \hline
\cite{huin2018energy}\cite{yang2016energy}\cite{soualah2017energy} & $\checkmark$ & $\checkmark$ & $\times$ & $\checkmark$ & $\checkmark$  \\ \hline
\cite{el2016energy} & $\checkmark$ & $\times$ & $\checkmark$ & $\times$ & $\checkmark$ \\ \hline
\cite{eramo2017migration} & $\checkmark$ & $\times$ & $\times$ & $\checkmark$ & $\checkmark$ \\ \hline
\cite{marotta2017energy} & $\checkmark$ & $\times$ & $\checkmark$ & $\checkmark$ & $\checkmark$  \\ \hline
\textbf{Our approach} & $\checkmark$ & $\checkmark$ & $\checkmark$ & $\checkmark$ & $\checkmark$ \\
\bottomrule
\end{tabular}
}\vspace{-1em}
\end{table}
\par Several studies investigated the VNF placement problem with different objectives (e.g., demand delay and cost). However, there are only a few studies that have focused on minimizing the energy/power consumption. El Khoury \textit{et al.}~\cite{el2016energy} formulated the problem to allocate and schedule traffic flows with deadlines to VNFs while minimizing the total PM power consumption in the network. Notably, in their scenario, VNF instances are already placed in the network. Kim \textit{et al.}~\cite{kim2017vnf} proposed an energy-aware VNF chain placement and reconfiguration algorithm based on Genetic Algorithm to minimize the PMs power consumption, considering QoS requirements in terms of latency. Pham \textit{et al.}~\cite{pham2017traffic} formulated an optimization problem to minimize the joint operational and network traffic cost. Their solution aimed at deploying as fewer number of PMs such that the communication cost between them was optimized. They used a sampling-based Markov approximation combined with matching theory to solve the combinatorial NP-hard problem in a quick way. Huin \textit{et al.}~\cite{huin2018energy} proposed an ILP to minimize the networking power consumption named \textit{GreenChains}. Yang \textit{et al.}~\cite{yang2016energy} studied VNF chain placement in datacenters. They provided an algorithm named \textit{Merge-RD} to save power in servers and network switches, and also reduce the transmission delay. Marotta \textit{et al.}~\cite{marotta2017energy} presented a joint resource and flow routing assignment mathematical model in order to minimize the power consumption of the hosting PMs and network devices. They used Robust Optimization theory to cope with data uncertainty in vEPC mobile networks. Actually, they defined a trade-off between power-efficiency, robustness, and the ability of the network to cover the data uncertainty. Eramo \textit{et al.} \cite{eramo2017migration} went a step further and proposed a server consolidation approach to achieve energy efficiency by turning off as many PMs as possible when the traffic density decreased. They formulated an optimization problem and proposed heuristics for online and offline cases, where the objective function is minimizing total PMs and VM migration energy consumption. Finally, Soualah \textit{et al.}~\cite{soualah2017energy} focused on minimizing the total PMs and network switches energy consumption. They formulated VNF placement and chaining problem by using the Monte Carlo Tree Search method and compared their solution to several other works.\par 
However, to the best of our knowledge, there is no work that tackled the VNF chain placement problem by jointly considering PMs and network power consumption, while satisfying end-to-end delay requirements. Furthermore, we apply the BI abstraction paradigm to enhance the VNF placement.
\section{System Model and Problem Formulation} \label{systemmodel} \vspace{.15cm}
We represent the network as a bidirectional graph $G(N,L)$, where $N$ is the set of nodes (network switches) and $L$ is the set of links in the network. In this design problem, every node is eligible to have a collocated PM that is able to host VNFs. Every PM has a set of resources $R$, which consist of: CPU, memory and storage. A set of demands $G$ is given, each demand $g$ been defined as:
{\small{
\begin{align}
    g(v_{s,g},v_{d,g},D_g,B_g,c_g)
    \label{eq_demand}
\end{align}
}}
where $v_{s,g}$ is the source node, $v_{d,g}$ is the destination node, $D_g$ is the end-to-end delay, $B_g$ is the required bandwidth capacity and $c_g = \{ f_1 \rightarrow f_2 \rightarrow ... \rightarrow f_{\mid c_g \mid }\}$ is the a set of ordered service functions, that is, the VNF chain $c_g$ of that demand.
VNF chains are modelled as a virtual network represented as a graph $G(V_g,E_g)$, where $V_g$ is the set of virtual nodes for demand $g$, and $E_g$ is the set of virtual links for demand $g$. The virtual nodes represent the functions (belonging to the given set of functions $F$) and endpoints of demand $g$, and the virtual links interconnect the virtual nodes following the order of that chain. In this way, $G(V_g,E_g)$ has to be mapped over $G(N,L)$ such that the demand $g$ requirements are guaranteed while minimizing the consumed power. The ingress and egress virtual nodes match the physical source and destination nodes in the substrate network.
\par A function $f$ can be reused by different demands as long as its maximum processing capacity [in Mb/s] $B_f$ is not surpassed. Otherwise, a new VNF should be placed in an active PM if it has enough resources. Finally, if active PMs do not have enough resources, a new PM must be activated. These three alternatives have different impact on the power consumption as introduced in the next section. 
\subsection{Power Model} \vspace{.1cm}
The following switch and PM power models are considered. The switch power consumption is defined as the sum of the power consumed regardless of the traffic load $P_{ss}$ and power proportional to the active ports~\cite{boru2015energy}, that is:
{\small{\begin{align}
    P_{sw}= P_{ss} + P_p N_p \label{eq_Pnet} 
\end{align}}}
where $P_p$ is the amount of power consumption of each active port, and $N_p$ is total number of active ports in a switch. \par
Considering that the active physical link $l_{i,j}\in L$ from node $i$ till node $j$ requires two ports, the total network power consumption can be expressed as: 
{\small{\begin{align}
    P_{net}^T = P_{ss} \sum_{i \in N} \nolimits y_i + 2P_p \sum_{ij \in L}  \nolimits l_{i,j} \label{eq_PnetT}
\end{align}}}
\noindent where $l_{i,j}$ is a binary variable that is equal to 1 if the physical link between node $i$ and $j$ is active. Notably, we assume a single link between each pair of nodes. Additionally, in the above equation, $y_i$ is a binary variable indicating if switch $i$ is active. \par
Since the CPU is the most significant power consumer in a server \cite{chase2001managing, chen2008energy, berl2010energy}, the power consumption model for the PM is based on its CPU utilization, which can be calculated using the following equation \cite{lee2012energy}:
{\small{\begin{equation}
    P_{pm} = P_{sm} + (P_{mm} - P_{sm}) \times \theta_{cpu} \label{eq_pm}
\end{equation}}}
\noindent where $P_{sm}$ corresponds to the static power consumption of the PM (i.e., the consumed power when its utilization is 0\%), $P_{mm}$ is the power consumption of the PM in its maximum CPU utilization and $\theta_{CPU}$ represents the CPU utilization. $\theta_{CPU}$ can be calculated as the ratio between the total CPU resources required and the available CPU resources available in the PM: $\theta_{CPU}=C_{f,r}/C^T_{i,r}$, where $r$ represents CPU resource type. Thus, the total power consumption of PMs in the network $P_{pm}^T$ can be expressed as follows: 
{\small{\begin{equation}
    P_{pm}^T = \sum_{i\in N} \nolimits \big(P_{sm} x_i + (P_{mm} - P_{sm}) \sum_{f \in F, r \in R} \nolimits \frac{C_{f,r}}{C^T_{i,r}} z_{i,f} \big) \label{eq_pmT}
\end{equation}}}
\noindent where $x_i$ is a binary variable indicating if the PM $i$ is active, and $z_{i,f}$ is an integer variable representing the number of instances of function $f$ placed on the PM at node $i$.
\subsection{ILP Formulation} \vspace{.1cm}
\label{sec:ILP}
Based on our previous work~\cite{vizarreta2017qos}, we mathematically formulate the joint power-aware VNF chain placement and routing problem using an ILP in order to obtain the optimal solution. The goal of this work is to optimally place VNF chains, such that overall power consumption of network switches, links, and PMs is minimized, while meeting end-to-end delay requirements. Therefore, the objective function is:
{\small{\begin{align}
    \text{Minimize} \hspace{.3cm} (P_{net}^T + P_{pm}^T). \label{eq5}
\end{align}}}
In addition to Eq. (\ref{eq_Pnet}) and (\ref{eq_pm}), the presented objective function must satisfy a number of constraints which are as follows:\par 
{\small {
\begin{align}
    &\sum_{f \in F}  \nolimits  C_{f,r} z_{i,f} \leq C^T_{i,r} \text{  ,}\hspace{.3cm} \forall i\in N,\forall r\in R, \label{eq6} \\
    & \sum_{g \in G} \nolimits  B_g u_{i,f,g} \leq B_f z_{i,f} \text{  ,}\hspace{.3cm} \forall i\in N ,\forall f\in F, \label{eq7}\\
    & \sum_{kl \in E_g} \nolimits \sum_{g \in G} \nolimits  B_g w_{ij,kl,g} \leq B_{i,j} \text{  ,}\hspace{.3cm} \forall i\in N,\forall j\in N, \label{eq8} \\
    & u_{i,f,g} \leq z_{i,f} \text{  ,}\hspace{.3cm} \forall i\in N,\forall f\in F,\forall g\in G, \label{eq9} \\
    & \sum_{i \in N}  \nolimits \sum_{f \in F} \nolimits  D_f u_{i,f,g} + \sum_{ij \in L} \nolimits  \sum_{kl \in E_g} \nolimits D_{i,j} w_{ij,kl,g} \leq D_g \nonumber \\
    &\text{, } \forall g\in G, \label{eq10} \\
    &\sum_{ij \in L} \nolimits  w_{ij,kl,g} - \sum_{ji \in L} \nolimits  w_{ji,kl,g} = u_{i,k,g} - u_{i,l,g} \hspace{.3cm} \label{eq11} \nonumber \\ 
    & \text{,}\forall i\in N,\forall k\in V_g,\forall l\in V_g, \forall g\in G, \\
    &u_{i,f,g} = 1 \hspace{.3cm} ,if\hspace{.3cm} i=f=v_{s,g}  \text{  ,}\hspace{.3cm} \forall g\in G, \label{eq12}\\
    & u_{i,f,g} = 1  \hspace{.3cm} ,if\hspace{.3cm} i=f=v_{d,g}  \text{  ,}\hspace{.3cm} \forall g\in G. \label{eq13}
\end{align}}}
Eq. (\ref{eq6}) indicates that the resources $C_{f,r}$ used by the functions placed in the PM that is connected to node $i$ must not exceed the available resources in it $C^T_{i,r}$. In Eq. (\ref{eq7}), the capacity requirements of all demands served by function $f$ in the PM in node $i$ is limited to the processing capacity of function $f$, where $u_{i,f,g}$ is a binary variable which equals to $1$ if function $f$ for demand $g$ is placed in PM $i$. Equation (\ref{eq8}) states that the bandwidth required by all demands served by link $(i,j)$ will not be larger than the capacity of the link $(i,j)$, noted as $B_{i,j}$, where $w_{ij,kl,g}$ is a binary variable that equals to $1$ if the physical link $(i,j)$ is used by the virtual link $(k,l)$ of demand $g$. In Eq. (\ref{eq9}), it is stated that demand and function mapping must match. In Eq. (\ref{eq10}), the total delay in the path, due to processing delay $D_f$ of all present functions $f$ in the PMs and the propagation delays $D_{i,j}$ in the physical links provided for demand $g$, is limited to the required end-to-end delay $D_g$ for demand $g$. The flow conservation law constraint is expressed in Eq. (\ref{eq11}), which states that for each network switch $i$, the difference of all outgoing and incoming physical links that are used for the virtual link between virtual nodes $k$ and $l$, and demand $g$ must be equal. Eq. (\ref{eq12}) and (\ref{eq13}) map the endpoints of the demand to the PMs in the substrate network.\par
Finally, let us introduce three indicator variables, defined in Eq.(\ref{eq14}), Eq. (\ref{eq15}), and Eq. (\ref{eq16}) respectively. These variables control the operation status (i.e., \textit{online} or \textit{offline}) of physical links, PMs, and network switches, respectively. Note that $\Psi$ is defined as a large positive number.
{\small{\begin{align}
    & \sum_{kl \in E_g}  \nolimits \sum_{g \in G} \nolimits  w_{ij,kl,g} \leq l_{i,j} \times \Psi \text{  ,}\hspace{.3cm} \forall i\in N,\forall j\in N, \label{eq14} \\
    & \sum_{j \in N} \nolimits  (l_{i,j} + l_{j,i}) \leq y_i \times \Psi \text{  ,}\hspace{.3cm} \forall i\in N, \label{eq15} \\
    &\sum_{f \in F}  \nolimits z_{i,f} \leq x_i \times \Psi \text{  ,}\hspace{.3cm} \forall i\in N. \label{eq16}
\end{align}}}
\begin{figure}[t]
\centering
\includegraphics[width=.8\linewidth]{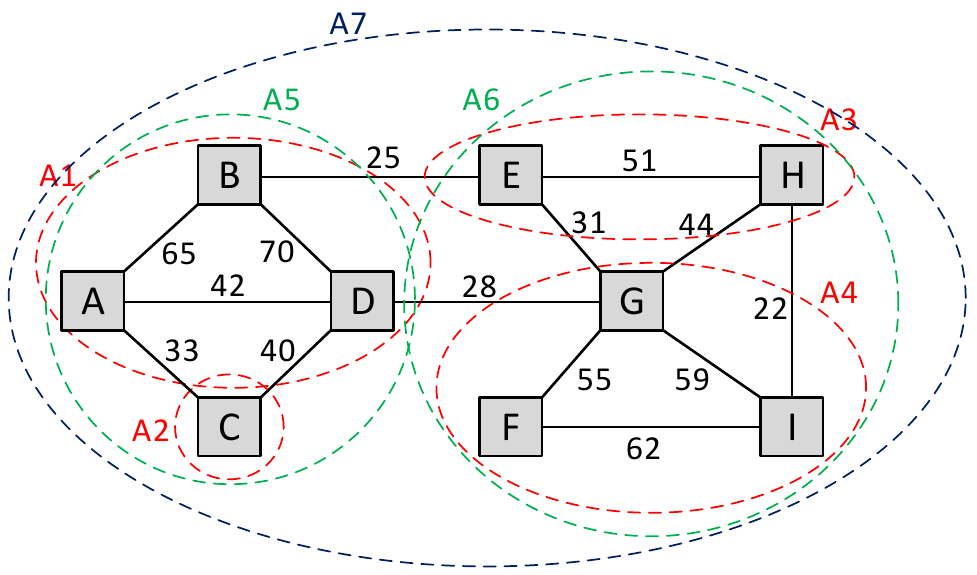}
\caption{An example of BIG, when $\beta$ values are 50, 40, and 30. A1-A4 are $50$-BIs, A5-A6 are $40$-BI, and A7 is $40$-BI. Link weights are their available bandwidth.}
\label{fig1}
\end{figure}
\section{Proposed Heuristic} \label{solution} \vspace{.15cm}
In order to cope with the ILP scalability issue, we present an effective heuristic approach, which consists of two main steps: 1) Network abstraction using the Blocking Island (BI) technique; and 2) Power-aware joint VNF chain placement and routing.
\subsection{Blocking Island Paradigm} \vspace{.1cm}
The BI technique, which is derived from Artificial Intelligence, is a resource abstraction method to represent the availability of resources (in this paper, bandwidth) in a graph \cite{frei1997simplifying}. The goal is to abstract the network into different sub-graphs (for different $\beta$ values). \par 
\textbf{$\beta$-BI Definition:}\cite{frei1997simplifying} A $\beta$-BI for node $x$ is the set of all nodes of the network that can be reached from $x$ using link(s) with at least $\beta$ available bandwidth.\par
Let us consider the network depicted in Fig. \ref{fig1} with 9 nodes and 14 links. $A1$ is the 50-BI for node $B$ (i.e., $\beta=50$) since all the nodes of $A1$ can reach $B$ and any other node of $A1$ with a path with at least 50 units of bandwidth. Since node $C$ can be connected with at most with 40 units of bandwidth, it does not belong to $A1$. The $\beta$-BI for a give node $x$ can be obtained by using a fast greedy algorithm, called $\beta$-Blocking Island Search (BIS) \cite{frei1997simplifying}, which its worst-case complexity is linear in $O(|L|)$ where $L$ is the set of links in the graph.

\par
\begin{figure}[t]
    \centering
    \subfloat[][]{\includegraphics[width=.4\linewidth]{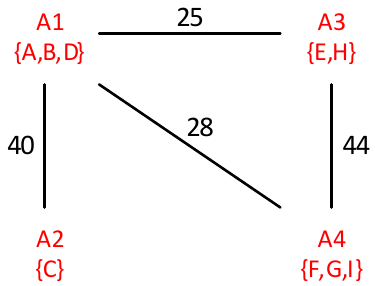}}
    \subfloat[][]{\includegraphics[width=.6\linewidth]{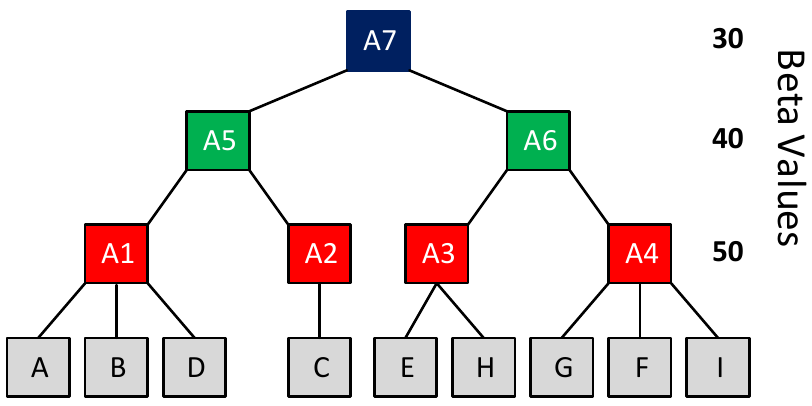}}
    \caption{(a) 50-BIG and (b) BIH tree from network in Fig. \ref{fig1} }
    \label{fig2}
\end{figure}

 For each $\beta$ value and the network graph, different $\beta$-BIs can be found, which can be represented as a $\beta$-BI Graph (BIG). For example, Fig.~\ref{fig2}(a) depicts the 50-BIG for the Fig. \ref{fig1} graph. The \textit{abstract nodes} are the different $\beta$-BIs and the \textit{abstract links} the network links interconnecting each pair of islands with the maximum bandwidth. 
Furthermore, for a given set of $\beta$ values $\mathscr{B} = \{\beta_1, \beta_2, ... , \beta_{|\mathscr{B}|}\}$, an overall network abstraction can be presented (so-called Blocking Island Hierarchy (BIH)). BIGs can be decomposed as a vertical tree structure in decreasing order of $\beta$s as shown in Fig. \ref{fig2}(b) for our example. This abstraction tree can reflect the real-time state of the available network bandwidth.\par

\subsection{BI-Based Heuristic} \vspace{.1cm}
For a given demand defined by Eq~\ref{eq_demand}, the BI paradigm first looks for $\beta$-BIs with $\beta \geq B_g$ containing both source and destination. If there are not any, the demand is rejected, since it can not be assigned, otherwise, the algorithm will proceed. Hence, the search space and time complexity decreases significantly. However, BIH should be updated every time a new demand is served by recomputing the BIs involved in that demand. This re-computation is performed in terms of merging or splitting operations which their complexities are only $O(n)$ or $O(l)$ \cite{frei1997simplifying}, where $n$ and $l$ denote the number of involved nodes and links on the new demand, respectively. Besides, experimental results show that despite updating costs, BI can greatly improve overall computation efficiency \cite{wang2015achieving}.
\label{sec:heuristic}
Since the VNF chain placement and routing problem is NP-complete, the ILP formulation presented in Section~\ref{sec:ILP} is not scalable. The BI paradigm is proposed as an efficient and fast solution. The heuristic algorithm consists of four main steps, which is presented in Algorithm \ref{alg2}. These steps are:\par
\noindent \textbf{Step 1:} Given the network graph $G(N,V)$ and a set of $\beta$ values $\mathscr{B} = \{\beta_1, \beta_2, ... , \beta_{|\mathscr{B}|}\}$, the BIH is built (line 1).\par
\noindent \textbf{Step 2:} For every demand defined as Eq. (\ref{eq_demand}), a BI is selected ($selectedBI$) that contains the source and destination and $\beta \geq B_g$) (line 3). In this study, two BI selection approaches have been considered:\par
\noindent \textbf{\textit{(i)} Highest $\beta$-BI (HBI):} In case several BI are available, the one with highest highest $\beta$ value in the BIH. By using this approach, the selected BI for hosting the request may contain less number of nodes. Thus, it can be expected that the probability of reusing VNFs decreases (i.e., more power consumption). Also, since HBI selects a BI with fewer nodes, fewer nodes should be checked for VNF chain placement. In this way, the time complexity is expected to reduce.\par
\noindent \textbf{\textit{(ii)} Lowest $\beta$-BI (LBI):} In case several BI are available, the BI with the lowest $\beta$ value in the BIH is selected. With LBI, the selected BI may contain more nodes than HBI. Hence, it is expected that the VNF reuse probability increases (i.e. less power consumption) and also incurs in more time complexity. The performance of these approaches is analysed in Section~\ref{evaluation}.\par
\begin{algorithm}[t]
\small
\caption{\small Power-Aware VNF Chain Placement and Routing}
\label{alg2}
\SetAlgoLined
\SetKwInOut{Input}{Input}
\SetKwInOut{Output}{Output}
\Input{Graph $G(N,V)$, Demands $(G)$, Beta Values ($\mathscr{B}$)}
\Output{Solution ($SOL = \{ sol_1,sol_2,...,sol_{|G|}\}$), where $sol_g$ consists of selected PM ($M$) and Routing($R$) for demand $g$}
\BlankLine
 BIList = ConstructBI($G,\mathscr{B}$)\;
 \ForEach {$g$ in $G$} {
    selectedBI = getBI($g$,BIList)\;
    MinCost = $\infty$  ;  $sol_g$ = \textit{null}\;
    \ForEach{$f$ in $c_g$}{
        M = \textit{null}; R = \textit{null}\;
        candidatePMs = getCandidatePMs($f$, selectedBI)\;
        \If{candidatePMs $== null$}{
            \textbf{break}\;
        }
        \ForEach{M in candidatePMs}{
            $R$ = calculateBestPath($g$, $M$, selectedBI)\;
            \If{R == null}{
                \textbf{continue}\;
        }
            solutionCost = calculatePowerCost($M$,$R$)\;
            \If{solutionCost $<$ MinCost}{
                    MinCost = solutionCost\;
                    $sol_g$.updatePM($M$)\;
                    $sol_g$.updateRouting($R$)\;
        }
        }
    }
    \If {$sol_g$ $\neq null$}{
        updateBI()\;
        SOL.append($sol_g$)\;
    }
 }
\textbf{return} SOL\;
\end{algorithm}
\begin{algorithm}[t]
\small
\caption{\small calculateBestPath}
\label{alg5}
\SetAlgoLined
\SetKwInOut{Input}{Input}
\SetKwInOut{Output}{Output}
\Input{Demand($g$), selectedPM, selectedBI}
\Output{Path between demand source($v_{s,g}$) and destination($v_{d,g}$) which passes through selectedPM satisfying delay($D_g$)}
\BlankLine
$\gamma = 1$; $\Omega = 0$ ; $\Delta_w$ = 0.25\;
\While{true}{
    \ForEach{Link in selectedBI}{
        Update \textit{edge\_weight} according to Eq.~(\ref{eq_weight})\;
    }
    $path_1$ = getShortestPath($v_{s,g}$, selectedPM)\;
    $path_2$ = getShortestPath(selectedPM, $v_{d,g}$)\;
    $path = path_1 + path_2$\;
    \eIf{getDelay(path) $<= D_g$}{
        \textbf{return} path\;
    }
    {
        $\gamma = \gamma - \Delta_w$\;
        $\Omega = \Omega + \Delta_w$\;
    }
    \If{$\gamma == 0$ or $\Omega == 1$}{
        \textbf{return} \textit{null}\;
    }
}
\end{algorithm}
\noindent \textbf{Step 3:} The joint power-aware VNF chain placement and routing algorithm tries to place the VNF instances and to find the route of this demand such that the total network and PM power is minimized. VNFs can be reused if they have enough capacity or they can be replicated at a new server, which can be active or should be turn it on with an extra power consumption. This algorithm consists of two main parts which are:\par
\noindent \textbf{Step 3.1:} VNF Placement: As it can be seen in Alg. \ref{alg2} lines 6-7, for every network function $f$ of the chain $c_g$, the PMs of $selectedBI$ that can host function $f$ are found with \textit{getCandidateList}. These $candidatePMs$ PMs can be \textit{(i)} PMs that already have an instance of function $f$with enough processing capacity to accept $g$, \textit{(ii)} PMs that are online and have enough available resources for creating a new instance of $f$, or \textit{(iii)} PMs that are offline. Note that the power cost of the option is increasing from \textit{(i)} to \textit{(iii)}.\par
\noindent \textbf{Step 3.2:} Power-aware routing: It finds the path from demand source $v_{s,g}$ to its destination $v_{d,g}$ containing intermediate node $candidatePM$. The proposed power-aware routing algorithm is presented in Alg. \ref{alg5}. This algorithm considers an edge weight \textit{edge\_weight}, which is a weighted sum of normalized values of consumed power \textit{edge\_power} and delay \textit{edge\_delay} of the edge. The normalization of \textit{edge\_power} and \textit{edge\_delay} is calculated based on the maximum possible power consumption and maximum link delay in the network, respectively. The \textit{edge\_Power} value of link $(i,j)$ is calculated as $(1-state(S_i)) \times \frac{1}{2} P_{ss} + (1-state(S_j)) \times \frac{1}{2}P_{ss}+ (1-state(l_{i,j})) \times 2P_p$ (see Eq.~(\ref{eq_Pnet})), where $state(S_i)$ and $state(L_{i,j})$ are equal to 1 if the status of switch $i$ and link $(i,j)$ is online, respectively, otherwise 0. Therefore, the expression of the edge's weight can be written as:
{\small{\begin{align}
    edge\_weight = (\gamma \times \widehat{edge\_power} + \Omega \times \widehat{edge\_delay}). \label{eq_weight}
\end{align}}}
The weighting parameters $\gamma$ and $\Omega$ are set to highlight the relevance of power over delay. The initial value for $\gamma$ and $\Omega$ is set to $1$ and $0$, but they will change gradually during the execution of the algorithm.\par
Once the edge weights are assigned, the algorithm tries to find a path from source to the $selectedPM$ and from the $selectedPM$ to the destination. The total path's delay is computed and compared with the demand's delay requirement ($D_g$). If the path's delay is compliant with the requirement (i.e., $\leq D_g$), the path is accepted. Otherwise, the edge weights are updated according to a given $\Delta_w$ value, which allows changing weights $\gamma$ and $\Omega$. This iterations continue until $\gamma = 0$ or $\Omega = 1$ (delay becomes the only priority). Actually, we use these weights to increase the demand acceptance rate by lowering the power-efficiency priority in each iteration. Indeed, there is a trade-off between power-efficiency and demand acceptance rate, since reusing network resources can increase the delay, which can lead to demand rejection. Nevertheless, the algorithm can be tuned by determining bounds for $\gamma$ and $\Omega$ and also the value of $\Delta_w$ for different objectives.\par

After $selectedPM$ is determined, the power consumption of the candidate solution can be calculated, which is the sum of the PM and network power (Eq.~(\ref{eq_pm}) and (\ref{eq_Pnet}), respectively). The solution with the minimum cost is kept as potential final solution for function $f$. Notably, if there are two solutions with the same cost, we select the PM which is the closest one to the source. This will leave more PMs to explore on the shortest-path. This procedure continues until all PMs in the $selectedPM$ list are examined.\par
\noindent \textbf{Step 4:} The affected BIs in the graph are updated (line 26) and add the selected PM ($M$) and routing ($R$) for demand g ($sol_g$) to the solution list ($SOL$). The same method applies to all demands in the demand set $G$ (starting with Step 1). Finally, the solution for all demands (i.e., $SOL$) is returned and the algorithm terminates.
\subsection{Betweenness Centrality-Based Heuristic} \vspace{.1cm}
In order to evaluate the advantages of using BI paradigm, a second heuristic algorithm based on the \textit{betweenness centrality (BC)} property of a graph has been developed. The BC of a node $v$\cite{freeman1977set} is defined as:
{\small{\begin{align}
    g(v) = \sum_{s \neq v \neq t} \nolimits \frac{\sigma_{st}(v)}{\sigma_{st}} \label{eq18}
\end{align}}}
where $\sigma_{st}$ is the total number of shortest paths from $s$ to $t$ and $\sigma_{st}(v)$ is the number of paths that pass through $v$. This BC-based algorithm first calculates the BC value for every node in the graph. Then, for each demand and its shortest path, the algorithm tries to place the required VNFs on the nodes with higher BC value. According to the BC definition, a higher BC value means higher number of shortest paths that are passing through that node $v$ and hence, higher probability of VNF reuse, which can lead to power-efficiency in a network.
\section{Performance Evaluation} \label{evaluation} \vspace{.15cm}
In this section, the proposed heuristics are compared with the ILP solution. The comparison is done in terms of the power consumption, end-to-end delay, acceptance rate, BI selection and runtime.
\begin{table}[t]
\setlength\belowcaptionskip{12pt}
\caption{Considered service function chains \cite{huin2018energy}}
\label{table3}
\resizebox{\columnwidth}{!}{
\begin{tabular}{ccccc}
\toprule
\textbf{Service Type} &\textbf{ VNF Chain} & \textbf{Bandwidth} & \textbf{Delay} & \textbf{\% Traffic} \\ \hline 
\textbf{Web Service} & NAT-FW-TM-WOC-IDPS & 100 kbps & 500 ms & 18.2\% \\ \hline
\textbf{VoIP} & NAT-FW-TM-FW-NAT & 64 kbps & 100 ms & 11.8\% \\ \hline
\textbf{Video Streaming} & NAT-FW-TM-VOC-IDPS & 4 Mbps & 100 ms & 69.9\% \\ \hline
\textbf{Online Gaming} & NAF-FW-VOC-WOC-IDPS & 50 kbps & 60 ms & 0.1\% \\
\bottomrule
\end{tabular}}\vspace{-1em}
\end{table}
The proposed ILP model has been implemented using the Gurobi Optimizer 5.0 solver in Python. The BI-based and BC-based heuristics have been implemented in Java environment. The simulations were executed on a machine equipped with Intel Core i7-6700 @3.40 GHz, 16 GB of RAM, running Windows 10 x64 OS.

The used network topology is the SNDLib \textit{Nobel Germany} with 17 nodes and 26 links \cite{orlowski2010sndlib}. The capacity of each physical link was set to 1 Gbps and their delay were determined proportional to their length considering optical fiber transmission. PMs are located at each node, each PM being able to host VNF instances. We assumed each PM is equipped with a 16-core processor. The power models for network and PM follow Eq. (\ref{eq_Pnet}) and Eq. (\ref{eq_pm}), respectively. The considered values are: switch power $P_{ss} = 130$W, link power $P_p = 1$W \cite{mahadevan2011energy}, static PM power $P_{sm}=150$W, and maximum PM power $P_{mm}=250$W \cite{beloglazov2012energy}.

The VNF chains described in \cite{huin2018energy} have been considered and summarized in Table~\ref{table3}. These chains include different service types with specific ordered set of VNFs and different requirements in terms of bandwidth and delay. Different function types have been considered: Network Address Translation (NAT), Firewall (FW), Traffic Monitor (TM), WAN Optimization Controller (WOC), Video Optimization Controller (VOC), and Intrusion Detection System (IDS). In addition, network traffic is divided to these service types by a specific percentage (e.g., 11.8\% for VoIP service type according to Table \ref{table3}). Using this information, different sets of demands can be generated. We uniformly distributed the source and destination nodes, and randomly assigned each demand to a service type based on the traffic percentages. All VNFs are assumed to have 200 Mb/s traffic processing capacity, and 10 ms of processing delay. \textcolor{black}{We used 10 ms to put stress on the ILP and heuristic algorithms, according to large end-to-end service delay requirements in Table \ref{table3}}. We also assumed that each function requires 4 CPU cores to be able to operate. Notably, static $\beta$ values as $\mathscr{B} = \{900,700,500,300\}$ and LBI BI selection approach have been considered for the first simulations. Each simulation has been executed 30 times so that mean value and standard deviation can be plotted in the figures. 

\begin{figure}[t]
\centering
\includegraphics[width=.93\linewidth]{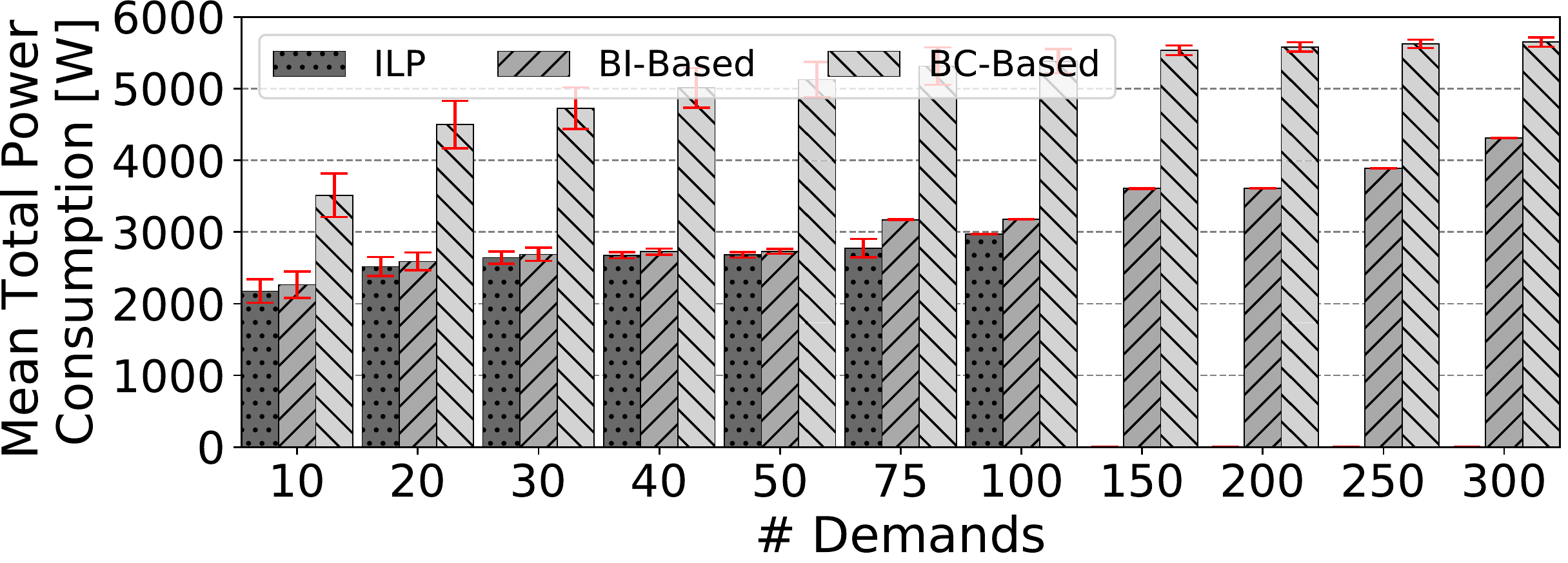} {\small{(a)}}
\includegraphics[width=.93\linewidth]{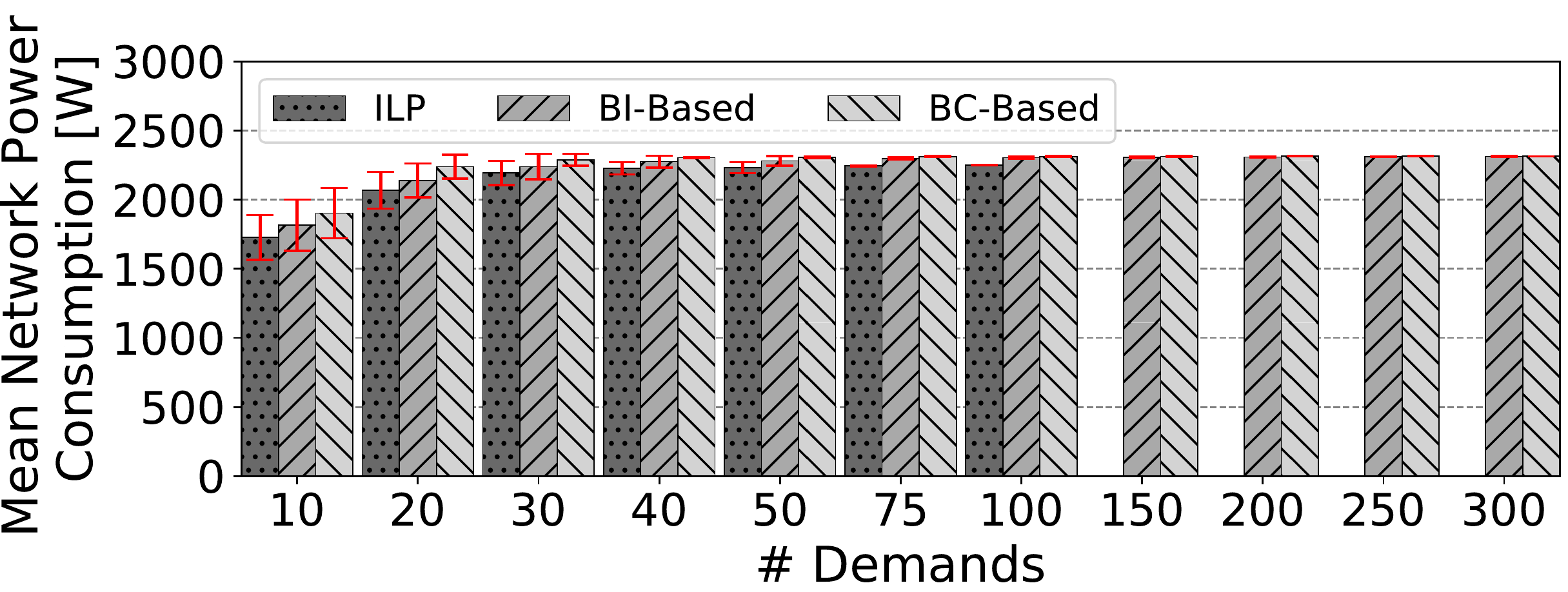} {\small{(b)}}
\includegraphics[width=.93\linewidth]{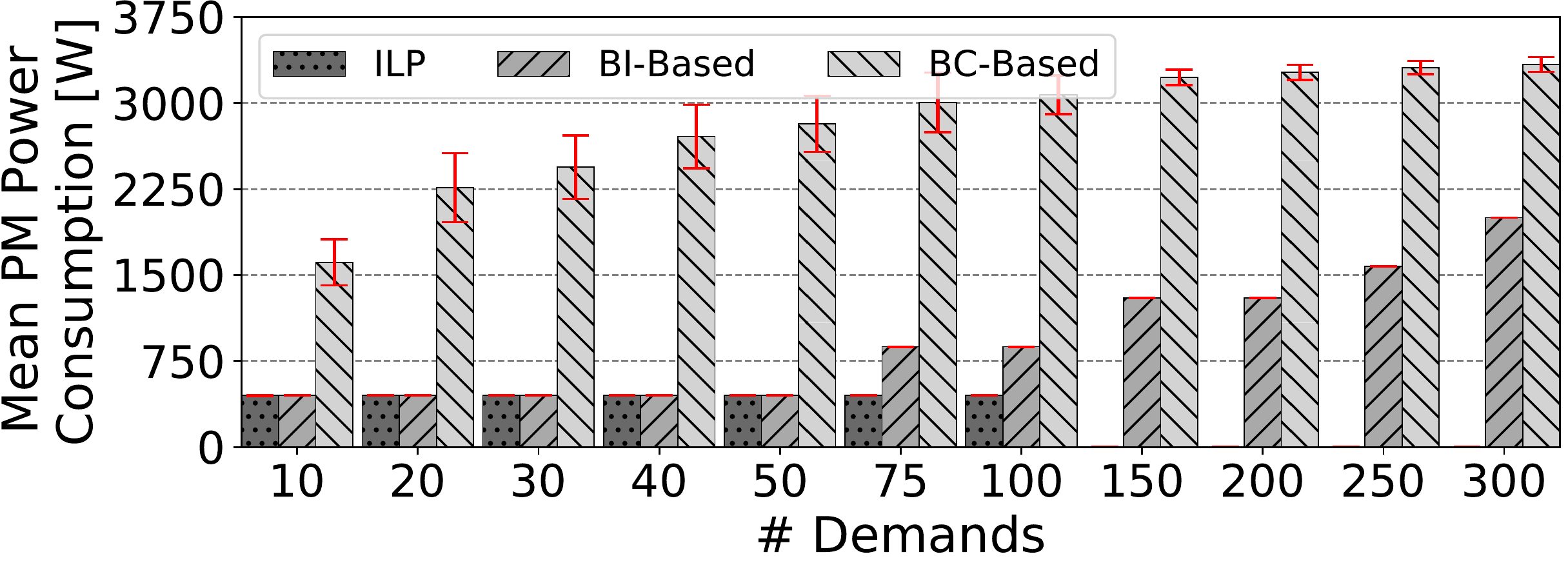}{\small{(c)}}
\caption{Total, network, and PM power consumption comparison}\label{fig:power}
\end{figure}

\begin{figure}[b]
\centering
\includegraphics[width=.93\linewidth]{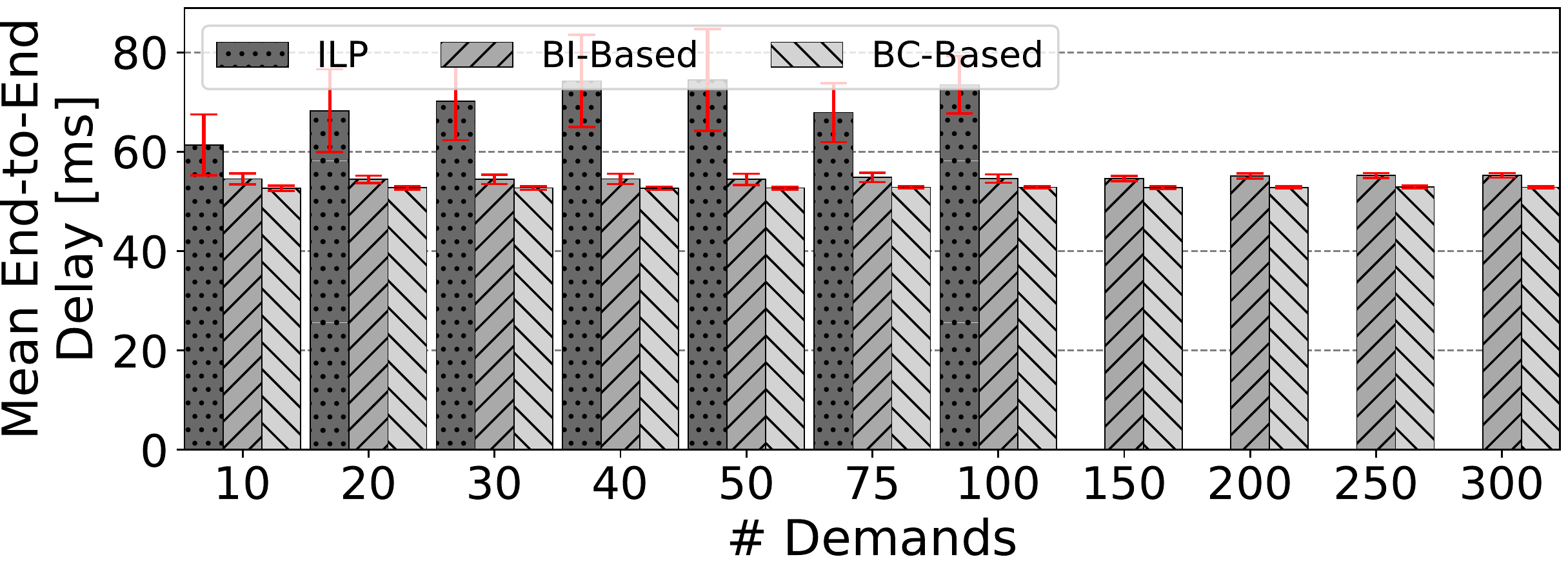}
\caption{Comparison of the average total end-to-end delay.}
\label{fig6}
\end{figure}

\par \noindent \textbf{A. Power Consumption:} The total power consumption results are compared in Fig.~\ref{fig:power}. It can be seen that the ILP can only solve small problem instances, while BI-based and BC-based heuristic are more scalable. The results show that the total power consumption achieved by the BI-based algorithm is closer to the optimal solution than the BC-based algorithm. Also, the mean PM power consumption in BC-based algorithm is significantly higher than the BI-based and the optimal solution. That is because it selects the shortest path, instead of using longer but more power efficient paths like BI-based and ILP. It can also be observed that the network power consumption of the BI-based outperforms the BC-based algorithm, while it is close to the optimal solution. Also, we note that for more than 100 demands, the two heuristics consume almost the same amount of power for operating the networking part. This is because the network is overloaded with the demands.
\par \noindent \textbf{B. End-to-End Delay:} Fig. \ref{fig6} presents the average total end-to-end delay for all demands and compares this parameter for the three proposed approaches. It can be seen that the delay obtained by the BC-based algorithm is always lower than the one achieved by the BI-based one. The reason is the BI-based algorithm may select the longer path when allocating a demand (compared with the BC-based algorithm, which takes the shortest path).
On the other hand, ILP model incurs higher delay than the two heuristics. This is because the ILP objective is to minimize total power consumption, so it reuses the resources as much as possible which adds delays due to a longer path. Thus, there is a trade-off between delay and power consumption in this scenario. 
\par
\begin{figure}[t]
\centering
\includegraphics[width=.93\linewidth]{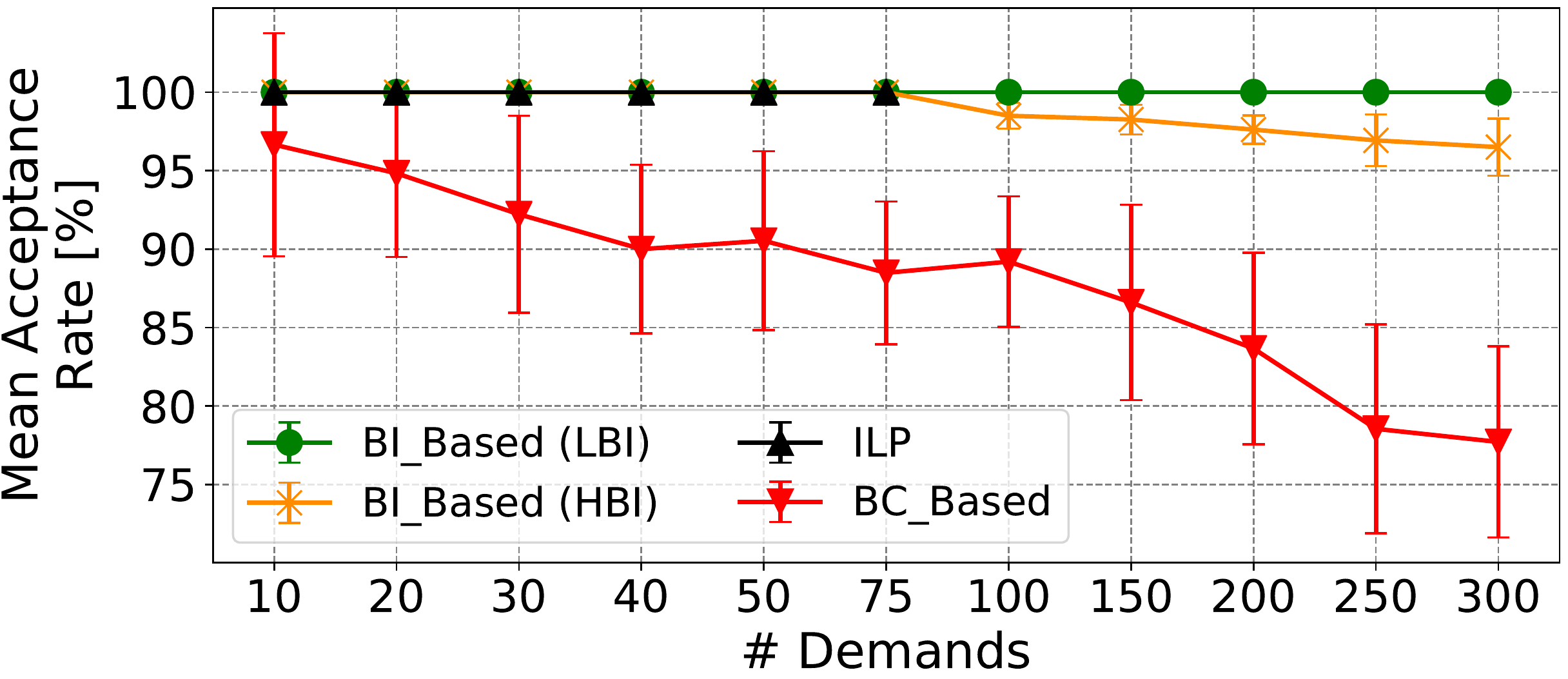}
\caption{Comparison of acceptance rates.}
\label{fig5}
\end{figure}

 \noindent \textbf{C. Acceptance Rate:} It is defined as the percentage of demands that have been hosted in the network. Fig. \ref{fig5} compares the acceptance rate for the ILP model, BI-based, and BC-based heuristics. In all the simulations with different sets of demands, ILP and BI-based approaches are able to serve all the demands. However, the BC-based algorithm cannot host all the demands is focusing only on the shortest path to be able to use the BC property. Hence, for the 300-demand case, it is able to accept only $80\%$ of the demands.
\par
\begin{figure}[b]
\centering
\includegraphics[width=.93\linewidth]{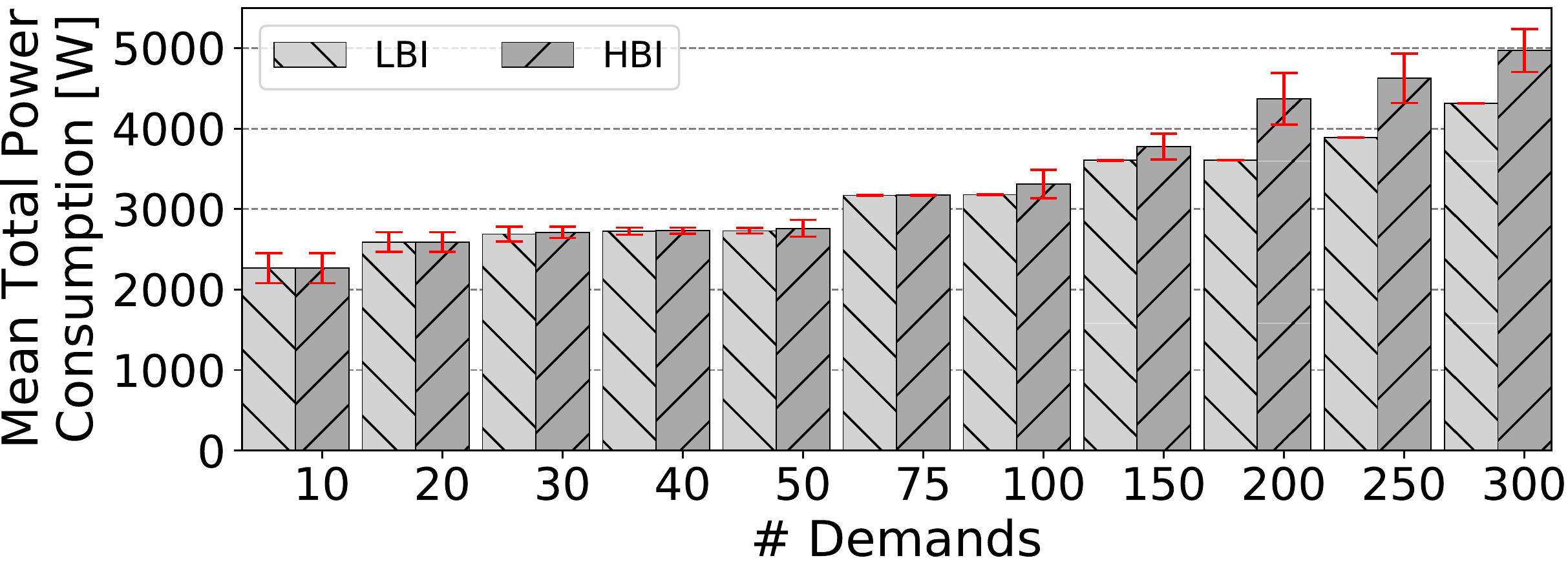}
\caption{Total power consumption for the HBI and LBI approaches.}
\label{fig7}
\end{figure}

 \noindent \textbf{D. BI Selection:} We compared the impact of the BI selection approaches LBI/HBI introduced in Section~\ref{sec:heuristic}, which select the lower/highest possible $\beta$ value, respectively. The power consumption for both methods is depicted in Fig. \ref{fig7}. It can be observed that LBI is more power-efficient than HBI. The reason is that, according to the BIH tree, getting a lower $\beta$ value means a larger BI with more potential PMs and reusable resources. This fact can also lead to different demand acceptance rate. As it is shown in Fig. \ref{fig5}, the mean acceptance rate for HBI approach is lower than LBI. However, since there are potentially fewer nodes in the HBI approach, its computational-efficiency is higher than LBI due to the reduced search space.

\vspace{.28cm} \noindent \textbf{E. Runtime:} Since the computation time is critical for the demand setup times, the computation time of the three solutions have been compared for different number of demands. As it can be seen in Table \ref{table4}, the ILP runtime (given in seconds) is significantly higher than the two heuristics. In fact, it grows exponentially with increasing number of demands compared with a linear increase of both heuristics. Additionally, it can be seen that the runtime of the BI-based algorithm is longer than the BC-based one due to the BI updates and more complex routing than the shortest path used by the BC-based algorithm. Notably, to make the solution time faster in ILP, we use $7\%$ solution gap for the case of 100 demands.

\section{Conclusion} \label{conc} \vspace{.15cm}
This paper addressed the problem of joint VNF chain placement and routing, meeting their end-to-end delay requirement, and also minimizing the overall power consumption by switches and PMs hosting the VNFs. We first proposed an ILP-based solution, and secondly two heuristics: \textit{(i)} Blocking Islands (BI) abstraction based algorithm, and \textit{(ii)} algorithm using the Betweenness Centrality (BC) property of a graph. The three solutions were compared in terms of the power consumption, the average demand's delay, the acceptance rate and the computation time. Our BI-based heuristic showed near-to-optimal results in terms of power consumption and an improvement in delay and computation time compared to the ILP, that is, the proposed BI-based heuristic is significantly faster and scales for large number of demands while reducing by 22\% the average demand's delay, with a penalty of consuming 6\% more power than the ILP solution. Furthermore, it was showed that choosing lower $\beta$ values (LBI) reduces the power consumption and increases the acceptance rate.
\begin{table}[t]
\setlength\belowcaptionskip{12pt}
\caption{Mean runtime for different approaches (seconds)}
\label{table4}
\resizebox{\columnwidth}{!}{
\begin{tabular}{cccccccc}
\toprule
\textbf{\#Demands} & \textbf{10} & \textbf{20} & \textbf{30} & \textbf{40} & \textbf{50} & \textbf{75} & \textbf{100} \\ \hline
\textbf{ILP} & 53.2 & 398.6 & 942.1 & 1742.1 & 2794.7 & 32269.3 & 18407 \\ \hline
\textbf{BI-Based} & 0.102 & 0.149 & 0.187 & 0.238 & 0.263 & 0.373 & 0.405 \\ \hline
\textbf{BC-Based} & 0.026 & 0.038 & 0.049 & 0.062 & 0.068 & 0.083 & 0.097 \\
\bottomrule
\end{tabular}
}\vspace{-1em}
\end{table}
\section{Acknowledgement}
This work was jointly supported under the Celtic-Plus sub-project SEcure Networking for a DATa center cloud in Europe (SENDATE)-PLANETS (Project ID 16KIS0261, Celtic-Plus project ID C2015/3-1) funded by the German Federal Ministry of Education and Research (BMBF), and the Celtic-Plus sub-project SENDATE-EXTEND funded by Vinnova (Celtic-Plus project ID C2015/3-3). This work was partly funded by the Swedish Research Council (VR) Framework project: Towards flexible and energy-efficient datacenter networks.

\bibliographystyle{unsrt}
\bibliography{main.bib}

\end{document}